**Article**  Open Access

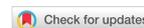

# Scalable solution chemical synthesis and comprehensive analysis of Bi$_2$Te$_3$ and Sb$_2$Te$_3$

Bejan Hamawandi[1,2,*], Parva Parsa[1], Inga Pudza[2], Kaspars Pudzs[2], Alexei Kuzmin[2], Sedat Ballikaya[3], Edmund Welter[4], Rafal Szukiewicz[5], Maciej Kuchowicz[5], Muhammet S. Toprak[1,*]

[1]Department of Applied Physics, KTH Royal Institute of Technology, Stockholm SE-106 91, Sweden.
[2]Institute of Solid State Physics, University of Latvia, Riga LV-1063, Latvia.
[3]Department of Engineering Science, Istanbul University - Cerrahpasa, Avcılar, Istanbul 34320, Turkey.
[4]Deutsches Elektronen-Synchrotron - A Research Centre of the Helmholtz Association, Hamburg D-22607, Germany.
[5]Institute of Experimental Physics, University of Wroclaw, Wroclaw 50-204, Poland.

**\*Correspondence to:** Prof. Muhammet S. Toprak, Department of Applied Physics, KTH Royal Institute of Technology, Stockholm SE-106 91, Sweden. E-mail: toprak@kth.se; Dr. Bejan Hamawandi, Institute of Solid State Physics, University of Latvia, Riga LV-1063, Latvia. E-mail: bejan.hamawandi@cfi.lu.lv

## Abstract

Thermoelectric (TE) materials can directly convert heat into electrical energy. However, they sustain costly production procedures and batch-to-batch performance variations. Therefore, developing scalable synthetic techniques for large-scale and reproducible quality TE materials is critical for advancing TE technology. This study developed a facile, high throughput, solution-chemical synthetic technique. Microwave-assisted thermolysis process, providing energy-efficient volumetric heating, was used for the synthesis of bismuth and antimony telluride (Bi$_2$Te$_3$, Sb$_2$Te$_3$). As-made materials were characterized using various techniques, including XRPD, SEM, TEM, XAS, and XPS. Detailed investigation of the local atomic structure of the synthesized Bi$_2$Te$_3$ and Sb$_2$Te$_3$ powder samples was conducted through synchrotron radiation XAS experiments. Radial distribution functions around the absorbing atoms were reconstructed using reverse Monte Carlo simulations, and effective force constants for the nearest and distant coordination shells were subsequently determined. The observed differences in the effective force constants support high anisotropy of the thermal conductivity in Bi$_2$Te$_3$ and Sb$_2$Te$_3$ in the directions along and across the quintuple layers in their crystallographic structure. The as-made materials were consolidated via Spark Plasma Sintering to evaluate thermal and electrical transport properties. The sintered TE materials exhibited low thermal conductivity, achieving the highest TE figure-of-merit values of 0.7 (573 K) and 0.9 (523 K) for n-type $Bi_2Te_3$ and p-type $Sb_2Te_3$, respectively, shifted significantly to the high-temperature region when compared to earlier reports, highlighting their potential for power generation applications. The scalable, energy- and time-efficient synthetic method developed, along with the demonstration of its potential for TE materials, opens the door for a wider application of these materials with minimal environmental impact.

## INTRODUCTION

The fast-changing and ever-increasing global energy demand has resulted in various fossil fuel-related environmental challenges, necessitating the development of renewable, affordable, and efficient energy supplies and energy-harvesting strategies[1-3]. Thermoelectric (TE) materials and devices can directly interconvert between heat and electrical power, which could contribute to energy generation, utilization, and environmental impact[3,4]. Based on the two fundamental phenomena, Seebeck and Peltier effects, TE devices may function as TE generators (TEGs) for harvesting heat energy or as coolers (TECs)[1,5]. TE materials-based technologies can be employed in various applications, including medical and wearable devices, low-power devices such as the Internet of Things (IoT), on-chip cooling, self-powering gadgets, solar cells, battery-free sensors, and space missions[1,5,6]. Thermal energy can be harvested from different low-grade heat sources, such as cars, buildings, electronic equipment, and the human body.

Unlike conventional electricity-generating technologies, TE devices function without the contribution of mechanical energy, making them reliable. They can operate under wide working temperature ranges without vibrations and noise, are easy to maintain, and have extended operational lifetime, among their attractive features[6-8]. However, low conversion efficiency, manufacturing challenges, complexity, and materials costs of this technology hinder its broad application[9,10]. The dimensionless TE figure of merit (ZT), expressed as $ZT = S^2\sigma T/\kappa$, describes the efficiency of a TE material, where $\sigma$ is the electrical conductivity, $S$ is the Seebeck coefficient, T is the absolute temperature, and $\kappa$ is the thermal conductivity[10-12]. The formula suggests that an ideal TE is a "phonon glass electron crystal", allowing charge carriers to flow freely (high $\sigma$) while maintaining a temperature difference across the material (low thermal conductivity)[10,13]. In addition, the material's band structure is critical for maintaining the thermal voltage (high $S$)[9,13]. Therefore, semiconductor materials, in general, are favorable as TE materials. However, since $S$, $\sigma$, and $\kappa$ have an inherent correlation in classical materials, improving the ZT has been challenging[14-16]. One of the crucial strategies to enhance the ZT is based on increasing phonon scattering through nanostructuring[17], producing nanoprecipitates[18], porosity design[19], alloying, and grain boundary engineering[20,21]. So far, numerous research efforts based on theoretical predictions and experimental explorations have shown that nanostructuring could considerably improve the performance of TE materials by lowering the negative correlation of TE transport characteristics. Due to a large density of interfaces (point defects/grain boundaries) in nanostructures, phonons are scattered more efficiently than electrons, since the mean free paths of heat carriers are longer than those of charge carriers. This results in a reduction in $\kappa$ (based on the phonon-blocking effect) while not considerably affecting the mobility of charge carriers[22,23]. Moreover, based on the quantum confinement phenomena, nanostructuring may widen the band gap and discretization in the electronic density of states, enhancing the $S$[13,17].

TE materials are categorized into three groups based on their operational temperature range: ambient, mid- and high-temperature. The band gap and chemical stability of these materials determine the feasibility of a given composition as a TE material. Most of the generated waste heat is attributed to temperatures below 200 °C, referred to as low-temperature waste heat. However, capturing and reusing low-grade waste heat is challenging since it has low energy density[24]. Binary and ternary metal chalcogenides perform well in this temperature range, with bismuth telluride ($Bi_2Te_3$) and antimony telluride ($Sb_2Te_3$) (and their alloys) being the two most popular n- and p-type TE materials with high ZT, operation around room temperature, effective up to 200 °C due to their specific band structure, high carrier mobility, and intrinsically low $\kappa$[23]. $Bi_2Te_3$ and $Sb_2Te_3$ are isostructural, with the structure along the *c*-axis composed of periodic quintuple layers (QL) with the sequence Te2-M-Te1-M-Te2, where M is Bi or Sb[25]. Two tellurium atoms (Te1 and Te2) are distinguished by their different local environments. Van der Waals (vdW) interactions between adjacent Te2 layers hold the QL stacked together and influence the material's mechanical, electronic, and TE properties.

TE materials have been synthesized through commonly used synthetic techniques, such as solid-state, gas-phase, and wet chemical routes. Each method has its constraints and advantages, leading to materials with different microstructures. Zone melting has long been the mainstream fabrication technique for the fabrication of commercial bismuth-telluride-based solid solutions[26]. Strategies for synthesizing nanostructured $Bi_2Te_3$ and $Sb_2Te_3$ include mechanical alloying[27], physical vapor deposition[28], chemical reduction[29], hydrothermal[3], solvothermal[30], and thermolysis[17,31,32] routes. Different morphologies of $Bi_2Te_3$ and $Sb_2Te_3$ (nanoparticles, nanowires, nanorods, and nanoplates) have been synthesized using these techniques. Over several decades, the ZT of these materials has improved. A detailed summary of processing time, costs, scalability, and TE properties associated with different methods can be found in earlier works[13,23,27]. However, a designed synthesis process is of great importance to produce the desired compositions and structures effectively and overcome the challenges that have restricted the commercialization of TE materials[12]. Considering this, some of the techniques listed above might be unsuitable for large-scale production due to their low purity, low yield, high energy cost, the possibility of contamination, and considerable batch-to-batch variations leading to limited reproducibility[3,10]. High-energy techniques such as zone melting, and melt alloying require typically high purity (5N) elemental powders, glove boxes and vacuum lines for handling the powders, and high-temperature furnaces. Bi-Sb tellurides grown by directional solidification, or zone-melting, method have poor mechanical properties because the basal plane in this rhombohedral structure is also the cleavage plane[33]. Among the wet-chemical synthetic techniques, the thermolysis method efficiently produces highly crystalline nanostructures with a narrow size distribution, high purity, and tunable morphology. Microwave (MW) heating can serve as the energy source for these synthesis techniques. Due to the effective volume heating, the MW-heating process is a fast, scalable, energy-efficient, and high-yield approach (with a high reproducibility) to obtain materials with well-defined chemical composition, high phase purity, and crystallinity. MW-assisted heating increases the overall kinetics of the reaction, hence resolving the problem of prolonged reaction time (from several hours/days to a few minutes) of the method. There are only a few reports[10,12,34-36] on the synthesis of $Bi_2Te_3$/$Sb_2Te_3$ using the MW-assisted thermolysis method, where the transport properties of the produced materials are presented (see Supplementary Table 1). Moreover, the number of studies examining detailed structural, microstructure analysis, surface morphology, and their effects on transport properties of $Bi_2Te_3$ and $Sb_2Te_3$ compounds is also limited. This work aims to establish an alternative facile and sustainable wet-chemical synthesis route for the large-scale synthesis of morphology-controlled, nanostructured n- and p-type TE materials ($Bi_2Te_3$ and $Sb_2Te_3$) through MW-assisted heating, with a focus on the time and energy efficiency, as well as the yield of the chemical process. The process uses readily available precursors and does not require glove box or high-temperature furnaces. We employed synchrotron radiation X-ray absorption spectroscopy (XAS) to gain insights into the local atomic environment and lattice dynamics around Bi, Sb, and Te atoms using an advanced data analysis technique such as reverse Monte Carlo (RMC) simulation. The as-made TE materials were consolidated into pellets via spark plasma sintering (SPS) to evaluate the characteristics of the electrical and thermal properties. Results indicate that the materials obtained exhibit a reasonable TE performance compared to many earlier reports, utilizing wet-chemical synthetic routes.

## EXPERIMENTAL

### Materials

Bismuth chloride (BiCl$_3$, 98% purity), Antimony chloride (SbCl$_3$, 99.95% purity), Tellurium powder (Te, 99.8% purity), Oleic acid (C$_{18}$H$_{34}$O$_2$), 1-Octadecene (C$_{18}$H$_{36}$, ODE), thioglycolic acid (C$_2$H$_4$O$_2$S, 98%, TGA), tri-butyl phosphine (C$_{12}$H$_{27}$P, 93.5%, TBP), acetone and isopropanol were all purchased from Sigma Aldrich (Stockholm, Sweden) and used as received without further purification. All chemicals were of analytical grade.

### Synthesis of nanostructured Bi$_2$Te$_3$ and Sb$_2$Te$_3$ powders

The synthesis was performed using a MW-assisted thermolysis method. To achieve the desired Bi$_2$Te$_3$ and Sb$_2$Te$_3$ compounds, stoichiometric amounts of each precursor were used to maintain a 2:3 molar ratio for Bi:Te and Sb:Te. The Te powder was first complexed with TBP (6 mL) by heating the mixture at 220 °C, using MW power of 400 W under constant stirring using a MW synthesizer (Biotage® Initiator+), until all the Te powder was fully dissolved. Meanwhile, in a separate vial, a precursor solution of Bismuth (or Antimony) was prepared by dissolving a stoichiometric amount of BiCl$_3$ (or SbCl$_3$) in Oleic acid under continuous stirring for 30 min. The solution was then transferred to a 100 mL Teflon vessel, where ODE (8 mL) and TGA (3 mL) were added. The reaction schematics are shown in Figure 1.

Eventually, the mixture was heated by MW (1800 Watt; flexiWAVE-Milestone, high-pressure multivessel rotor) to the reaction temperature of 220 °C with 4 min ramp time and 2 min dwell time (see Supplementary Figure 1A). Upon cooling down the solution to room temperature, the synthesized powder could be separated easily from the reaction mixture by simple decantation, as displayed in Supplementary Figure 1B. The products were washed once with acetone and twice with isopropanol, then dried in a vacuum oven at 80 °C for 3 h. Large-scale synthesis has been achieved by performing the same reaction in parallel in four reactors in a single run. The MW-reactor system, reaction profile and multivessel reaction possibilities are presented in Supplementary Figure 2.

### Consolidation using spark plasma sintering

The powders obtained by the MW-assisted thermolysis method were loaded into a graphite die (Φ 15 mm), with top and bottom graphite punches, and sintered by Spark Plasma Sintering (SPS, Dr. Sinter 825, Fuji Electronic Industrial Co. Ltd., Tokyo, Japan) to form nanostructured pellets. The sintering process was performed at 400 °C under 70 MPa with a heating rate of 30 °C/min and a holding time of 5 min for both samples. During the cooling process from 400 °C to room temperature, the load was reduced from 70 to 0 MPa. Finally, the pellet was polished to achieve a smooth and decontaminated (from residues of the graphite die and punches) surface for further analysis. The relative density of the compacted samples was about 78%.

### Structural, morphological, and surface characterization

*X-ray powder diffraction analysis*

X-ray powder diffraction (XRPD) analysis was accomplished using a Philips PANalytical X'Pert Pro Powder Diffractometer, equipped with copper anode (Cu-K$\alpha_1$ radiation, λ = 1.5406 Å) to identify the crystalline phases present, and the changes in the structure due to sintering. The analysis was done in the continuous scan mode, with a step size (2θ) of 0.24°, and 0.04°/s as the scan speed. The crystalline phases of the samples were determined via the High-score Pro software.

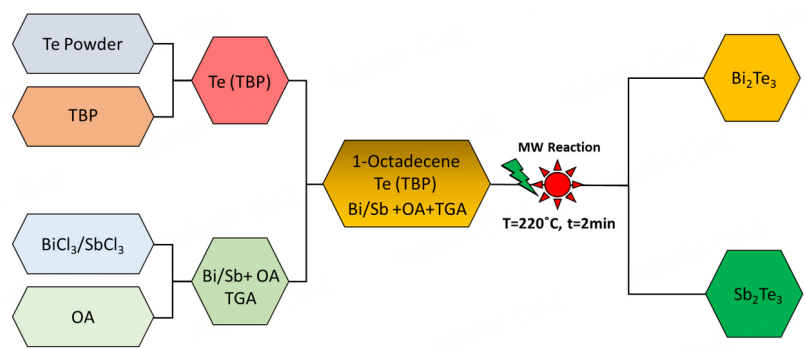

**Figure 1.** Schematics of the MW-assisted thermolysis synthesis of $Bi_2Te_3$ and $Sb_2Te_3$.

The particle size and morphology of the sample were studied using scanning electron microscopy (SEM, FEI Nova 200, Hillsboro, OR, USA). The SEM samples were fixed on the holder using graphene ink or copper tape, to avoid charging and enhance the image resolution. Transmission electron microscopy (TEM, JEM-2100F, 200 kV, JEOL Ltd. Tokyo, Japan) analysis was performed on the as-prepared nanopowders to investigate the crystallinity of the samples. Samples were made by drop casting a 200 μL aliquot of nanoparticle suspension in isopropanol on copper TEM grids and drying them in air.

*X-ray absorption spectroscopy analysis*
Temperature-dependent X-ray absorption spectroscopy (XAS) measurements of $Bi_2Te_3$ and $Sb_2Te_3$ samples were performed at DESY PETRA III P65 Applied XAS beamline[37]. The PETRA III storage ring operated at $E$ = 6.08 GeV and current $I$ = 100 mA in top-up 40 bunch mode. The harmonic rejection was achieved by uncoated (Bi $L_3$-edge) and Pt-coated (Sb and Te K-edges) silicon (Si) plane mirrors. Fixed-exit double crystal monochromators Si(111) and Si(311) were used. The X-ray absorption spectra were collected in transmission mode at the Bi $L_3$-edge (13,419 eV), Sb K-edge (30,491 eV), and Te K-edge (31,814 eV) using two ionization chambers. The samples were prepared from powders mixed with cellulose and pressed into pellets. Temperature control was achieved using the continuous-flow liquid He cryostat (SuperTran-VP, Janis Research Company, LLC) in the temperature range from 10 to 300 K.

*X-ray photoelectron spectroscopy analysis*
Surface chemical composition of the synthesized materials has been examined using the X-ray photoelectron spectroscopy (XPS) technique. The non-monochromatized X-ray source with a Mg anode lamp (Mg Kα line, 1,253.6 eV) was used in all analyses. High-resolution photoelectron energy spectra were recorded with Auger Electron Spectroscopy (AES)/XPS system EA10 (Leybold-Heraeus GmbH, Cologne, Germany) at room temperature, under the pressure less than $5 \times 10^{-8}$ mbar. The spectrometer energy scale was calibrated using Au 4f, Ag 3d, and Cu 2p lines, and all acquired spectra were calibrated to the adventitious carbon C 1s line at 285 eV. The overall resolution of the spectrometer during the measurements was estimated as 0.96 eV, using the full width of half maximum (FWHM) of the Ag $3d_{5/2}$ line. The data acquisition was done using WSpectra version 8 (R. Unwin, 2001) software, and data processing (deconvolution) was performed using CasaXPS software version 2.3 (Casa Software Ltd., Teignmouth, U.K.). A wide-range energy scan was collected to determine surface chemical composition, based on which the high-resolution scans of elemental emission lines were performed. The atomic concentration in the sample was determined through XPS spectral analysis, considering the presence of each element.

*Electrical and thermal transport property evaluations*

The TE transport measurements were conducted on SPS compressed pellets. The total thermal conductivity $\kappa_{tot}$ was estimated using $\kappa_{tot} = C_p \cdot \alpha \cdot \rho$, where $C_p$ is the specific heat capacity, $\alpha$ is thermal diffusivity, and $\rho$ is the density of mass. $C_p$ measurements were performed via Differential Scanning Calorimetry (DSC, PT1000 Linseis). The $\alpha$ was measured on 2 mm thick disk-shaped samples (see Supplementary Figure 3 for sample geometry) using the laser flash analysis (LFA 1000, Linseis) system. σ and *S* measurements were performed simultaneously using the commercial ZEM3 ULVAC-RIKO system. A percentage error of 5%-10% is possible in calculating the thermal conductivity of $Bi_2Te_3$ and $Sb_2Te_3$ pellets using the laser flash method and can vary due to several factors, including sample quality and measurement precision[38,39].

*Data analysis by reverse Monte Carlo simulations*

Experimental extended X-ray absorption fine structure (EXAFS) spectra were extracted from X-ray absorption spectra using the XAESA code[40], following the conventional procedure[41]. The analysis of EXAFS spectra employed the RMC method based on the evolutionary algorithm (RMC/EA) implemented in the EvAX code[42,43]. The RMC method is an atomistic simulation approach that iteratively minimizes the difference between experimental and calculated EXAFS spectra, making random atomic displacements within the structure model of the material.

In this study, an initial structure model was constructed for each temperature point based on lattice parameters from the neutron diffraction data[44]. A supercell (4a × 4b × 3c) containing 720 atoms with periodic boundary conditions was used for calculations. The RMC/EA calculations were performed for 32 atomic configurations simultaneously. At each iteration, all atoms in the supercell were randomly displaced with a maximum allowed shift of 0.4 Å. The configuration-averaged EXAFS spectra were calculated using the ab initio self-consistent real-space multiple-scattering (MS) FEFF8.5L code[45,46], considering contributions from single, double, and triple scatterings. The complex energy-dependent exchange-correlation Hedin-Lundqvist potential was employed to account for inelastic effects[47]. The amplitude scaling parameter is set to 1. Good agreement between experimental and simulated data was achieved in both *k*- and *R*-spaces simultaneously for two absorption edges (Sb/Bi and Te). This was demonstrated by comparing the Morlet wavelet transform (WT) of the respective EXAFS[48]. RMC/EA calculations for $Bi_2Te_3$ and $Sb_2Te_3$ were performed in the *k*-space range of 3.0-15.4 Å$^{-1}$ and the *R*-space range from 2 to 7.0-7.6 Å for both (Sb/Bi and Te) metal edges. Note that such a wide range in *R*-space allows one to determine the partial radial distribution functions (RDFs) $g(r)$ of up to about 8 Å. The convergence of each RMC simulation was achieved after several thousand iterations. Five RMC/EA simulations with different sequences of pseudo-random numbers were performed for each experimental data set. As a result of simulations, 3D structure models of the tellurides were obtained at each temperature point and used to calculate the partial RDFs $g(r)$ and the average interatomic distances and the mean-square relative displacement (MSRD) factors, also known as the Debye-Waller factors. The temperature dependencies of the MSRD factors for the nearest and distant coordination shells were further utilized to estimate the effective force constants (*f*).

## RESULTS AND DISCUSSION

### Structural analysis

Through MW-assisted thermolysis, it has been possible to synthesize nanostructured materials in an energy and time-efficient manner. The synthesis process took about 6 min, with a reaction yield of 98%. Structural analysis was performed using XRPD on as-synthesized powders, and after their sintering using SPS. The diffraction patterns of as-made and sintered samples are presented in Figure 2, where the diffraction patterns are normalized for the most intense diffraction indexed to the (015) plane. For the sintered pellet samples, the XRPD analysis was conducted on the surface perpendicular to the sintering direction. The crystalline phases are identified as $Sb_2Te_3$ (ICDD: 96-900-7591) and $Bi_2Te_3$ (ICDD: 01-085-0439), both exhibiting a rhombohedral crystal structure. The Bragg diffraction peaks have been indexed with the appropriate Miller indices, as displayed on the diffraction patterns in Figure 2. When comparing the sintered samples with as-made samples, it is evident that sintered samples have more intense Bragg diffraction peaks corresponding to the same Miller indices of (006), (1010), (0015), and (1016), indicating a higher crystallinity. Following sintering, a significant increase in the relative intensity of the diffraction peaks with the (00*l*) index is observed, in comparison to the dominant (015) peak, indicating a high level of texturing in both samples. The ratio of intensity of the (006) peak to (015) peak can be used to assess the degree of texturing. After the SPS process, the *c*-axis of the grains aligned parallel to the pressing direction, leading to a greater degree of texturing in the case of $Bi_2Te_3$.

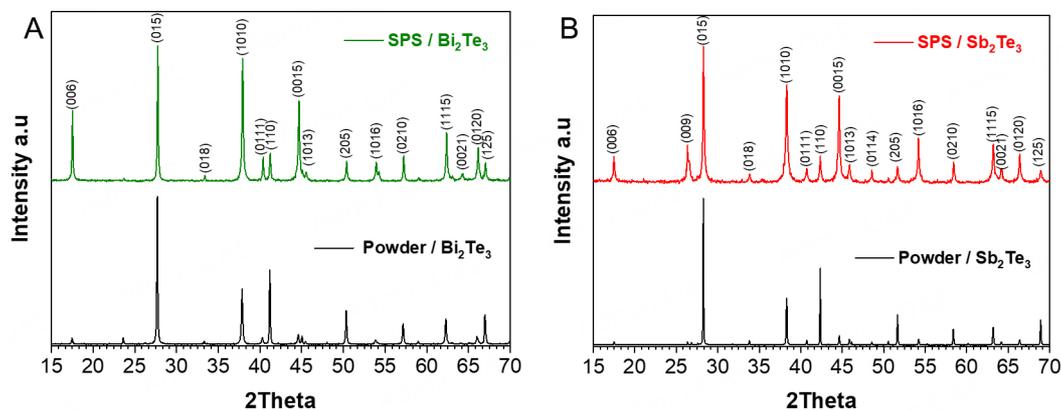

**Figure 2.** X-ray powder diffraction (XRPD) patterns of as-synthesized and sintered $Bi_2Te_3$ (A) and $Sb_2Te_3$ (B) samples. The major crystalline phases are indexed to $Bi_2Te_3$ (ICDD: 01-085-0439) and $Sb_2Te_3$ (ICDD: 96-900-7591), with rhombohedral crystal structure.

**Microstructure analysis**

Microstructure and morphology analyses for as-made powder and sintered samples as pellets have been performed using SEM. Micrographs at different magnifications for $Bi_2Te_3$ and $Sb_2Te_3$ samples are given in Figure 3A and B. As-made nanoparticles display hexagonal platelet morphology for both samples, where $Sb_2Te_3$ platelets were observed to be thinner than those of $Bi_2Te_3$. In general, nanoparticles of $Bi_2Te_3$ are smaller than those of $Sb_2Te_3$, even though they are made with the same route, with significantly different lateral dimensions, in the range of 20-500 nm for $Bi_2Te_3$ and 20-4,000 nm for the $Sb_2Te_3$ sample. Microstructure analyses were earlier reported for hydrothermal and polyol routes, suggesting different kinetics of nucleation and growth of $Bi_2Te_3$ and $Sb_2Te_3$[3,49], where microstructure control has been achieved by specific additives. The platelet morphology in the thermolysis process has been achieved by using TGA, which stabilizes the c-axis, enhancing the growth of the nanocrystals in the lateral (x-y) plane.

A more detailed examination was conducted using TEM analysis, and a few micrographs are presented in Figure 3C and D. The platelets show an almost perfectly single crystalline nature. Some platelets are aligned sideways, revealing the lamella-like structure along the *c*-axis, which display a thickness of about 1 nm. These five atomic layers (Te-Bi-Te-Bi-Te) are the QLs held together by the vdW forces.

Pellets of consolidated nanostructured materials were prepared using the SPS technique. A few SEM micrographs of consolidated samples are presented in Figure 4A-C for $Bi_2Te_3$ and Figure 4D-F for $Sb_2Te_3$ samples at different magnifications. Both samples show laterally growing grains and apparent stacking of hexagonal platelets after the sintering process, perpendicular to the c-axis, which reveals a high level of texturing. This was also observed in XRPD [Figure 2], with an increase in the intensity ratio of (006) and (015) diffraction peaks ($I_{006}/I_{015}$), especially for the $Bi_2Te_3$ sample.

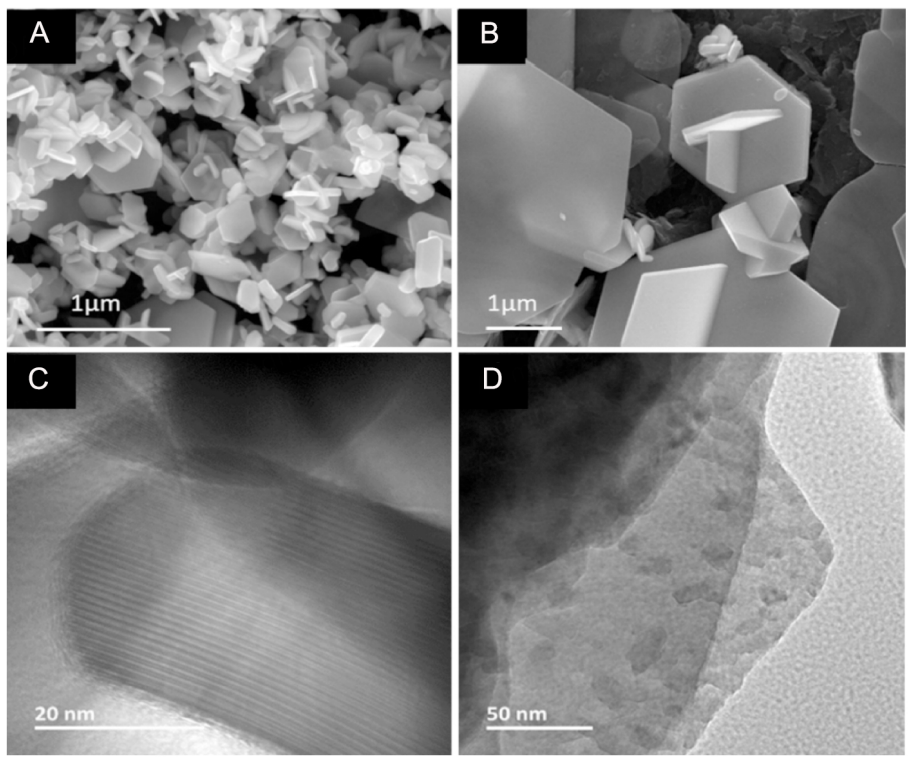

**Figure 3.** SEM and TEM micrographs of (A and C) as-made $Bi_2Te_3$ and (B and D) $Sb_2Te_3$ samples synthesized through MW-assisted thermolysis.

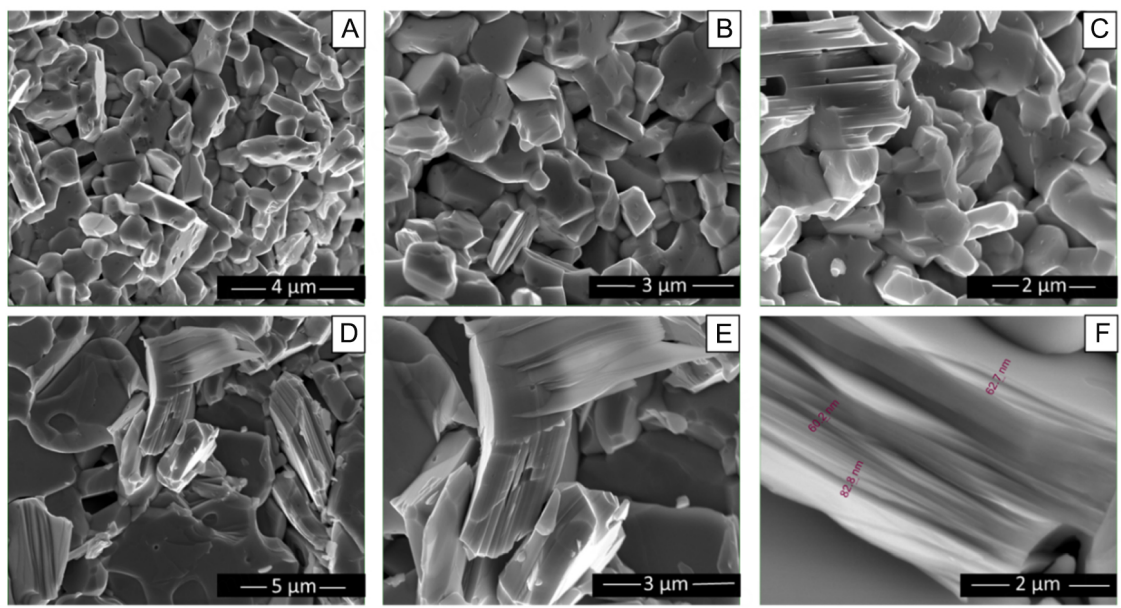

**Figure 4.** SEM micrographs, at different magnifications, of SPS sintered $Bi_2Te_3$ (A-C) and $Sb_2Te_3$ (D-F) samples, synthesized through MW-assisted thermolysis.

**XPS analysis**
XPS analysis has been used to investigate the surface chemical composition of $Bi_2Te_3$ and $Sb_2Te_3$ samples. A summary of the analysis results is presented in Table 1 and graphically in Figure 5, showing that all analyzed surfaces contain a specific amount of carbon and other elements - as listed in the table. The percentages of individual carbon bonds in the two samples are comparable: C-C bond (~60%), C-O (~22%), C-O=C (~10%), and carbonate (~8%) [Figure 5C and F]. In the case of tellurium, metallic Te residues and mixtures of $TeO_x$ ($TeO_2$ and $TeO_3$) or $Te(OH)_6$ were detected [Figure 5B and E]. Bi was present on the surface of $Bi_2Te_3$ only in the form of $Bi_2O_3$ oxide [Figure 5A]. In the case of Sb, only the Sb $3d_{3/2}$ area was quantitatively analyzed. Figure 5D shows an example of deconvolution that can be performed to separate the Sb $3d_{5/2}$ area from the O 1s region. Using this approach, the residues of individual Sb compounds were identified.

We have earlier reported on the surface chemistry of $Bi_2Te_3$ samples synthesized through hydrothermal and polyol routes[3]. A quick comparison of the surface chemistry of as-made $Bi_2Te_3$ through thermolysis with similar materials obtained through wet-chemical synthesis would help understand the process of inherited minute differences in surface speciation. Nanoparticles synthesized through the thermolysis route reveal a much higher content of organics (70.44% C) when compared to polyol (39.6% C) and hydrothermal (61.53% C) samples. Oxygen content (23.5%) is also higher than the polyol (22.29%) and hydrothermal (16.24%) samples. Another striking difference is observed in the speciation of Te, where the thermolysis sample possesses a higher amount of Te in oxide and hydroxide phases, with the presence of a significant amount of $TeO_2$, $TeO_3$, and $Te(OH)_6$ phases, of which $TeO_3$ and $Te(OH)_6$ phases were not observed in the polyol and hydrothermal samples.

**Local atomic structure and dynamics**
The local atomic structure of the as-synthesized $Bi_2Te_3$ and $Sb_2Te_3$ samples was studied using XAS. Figure 6 displays the experimental EXAFS spectra $\chi(k)k^2$ of both samples and their corresponding Fourier transforms (FTs) at the Bi $L_3$-edge and Sb/Te K-edges. At temperatures close to 300 K, the amplitude of EXAFS spectra is significantly reduced due to the large thermal disorder. Cooling down to 10 K significantly suppresses the thermal contribution that results in higher EXAFS amplitude and more pronounced FT peaks, revealing rich information on the local structure up to at least 8 Å. While the analysis of distant coordination shells is a challenging task using traditional methods, the RMC/EA method[43,50] provides a robust approach in this case. However, data analysis should be conducted cautiously, especially at elevated temperatures, where the experimental EXAFS spectra are strongly damped due to thermal disorder.

The RMC/EA simulations resulted in good agreement between experimental and calculated EXAFS spectra in both *k*- and *R*-spaces at all temperatures for $Sb_2Te_3$ [Supplementary Figure 4A] and $Bi_2Te_3$ [Supplementary Figure 4A and B]. It is important to note that a single structural model was derived for each sample by optimizing the agreement between the Morlet WTs [Supplementary Figure 4C and D] of the experimental and calculated EXAFS spectra simultaneously at two absorption edges (Sb/Bi and Te). In all RMC/EA simulations, the structural model aligns with the rhombohedral crystal structure with the space group $R\bar{3}m$[44]. At elevated temperatures, contributions from the nearest coordination shells dominate EXAFS, FT, and WT signals. However, at lower temperatures, contributions from outer shells become noteworthy. Indeed, at 10 K, the FT peaks at about 4 Å exhibit the highest intensity for both Sb and Te K-edges in $Sb_2Te_3$ and correspond to Sb-Sb and Te-Te scattering paths, respectively. Additionally, among many other structural peaks, a substantial contribution is apparent at about 6 Å (also seen as bright spots in WT [Supplementary Figure 4C]). Also, in the case of $Bi_2Te_3$, the layered structure gives rise to several distinguishable structural peaks, which are well observed at low temperatures [Supplementary Figure 4B and D].

**Table 1. XPS peak fitting for C 1s, Bi 4f, Te 3d, Sb 3d, and O 1s regions for $Bi_2Te_3$ and $Sb_2Te_3$ samples synthesized through MW-assisted thermolysis**

| Sample | Binding energy [eV] | Fraction | | Assignment |
|---|---|---|---|---|
| $Bi_2Te_3$ | 285 | 70.44 | C | C-C, C=O, C-O, carbonate |
| | 159 | 2.51 | Bi-O ($Bi_2O_3$) | |
| | 573 | 1.03 | Te | |
| | 575 | 0.88 | TeOx and $TeO_2$ | |
| | 577.5 | 2.09 | $TeO_3$, $Te(OH)_6$ | |
| | 530 | 23.05 | O | |
| $Sb_2Te_3$ | 285 | 72.81 | C | C-C, C=O, C-O, carbonate |
| | 528.3 | 0.23 | Sb | |
| | 530 | 1.86 | Sb-Ox | |
| | 573 | 0.15 | Te | |
| | 575 | 0.8 | TeOx and $TeO_2$ | |
| | 577.5 | 1.08 | $TeO_3$, $Te(OH)_6$ | |
| | 530 | 23.07 | O | |

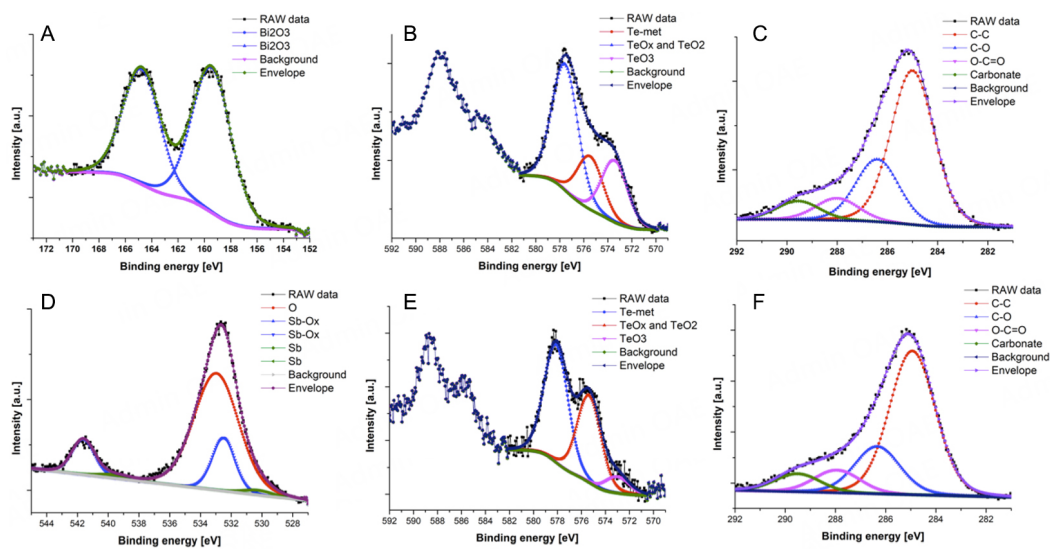

**Figure 5.** X-ray photoelectron spectroscopy (XPS). (A) Bi 4f, (B) Te 3d, (C) C 1s spectra of $Bi_2Te_3$; (D) Sb 3d, (E) Te 3d, and (F) C 1s spectra of $Sb_2Te_3$ samples synthesized through MW-assisted thermolysis.

The atomic coordinates obtained in the RMC/EA simulations were used to calculate partial RDFs $g(r)$ for Bi-Te, Bi-Bi, Sb-Te, Sb-Sb, and Te-Te atom pairs. The temperature dependence of RDFs $g(r)$ is shown in Figure 7. The RDFs were evaluated as an average of five independent RMC/EA simulations to improve statistics. The first peak in RDFs $g(r)$ corresponds to Sb(Bi)-Te2 and Sb(Bi)-Te1 atom pairs. The two contributions are split at 10 K. However, they merge, forming a broad, asymmetric peak at elevated temperatures.

Atomic coordinates were also used to calculate the average interatomic distances and the MSRD factors. The MSRDs were estimated using the median absolute deviation (MAD) function as described earlier[51]. The average distance within the first coordination shell [r1(M-Te2)] remains unchanged within error bars throughout the entire temperature range, measuring about 2.99 Å in $Sb_2Te_3$ and 3.07 Å in $Bi_2Te_3$. The average distance for the second coordination shell [r2(M-Te1)] increases from about 3.15 to 3.18 Å in $Sb_2Te_3$ and from about 3.24 to 3.26 Å in $Bi_2Te_3$ when heated from 10 up to 300 K. The Bi-Te interatomic distances are larger relative to Sb-Te due to differences in the radii of Bi and Sb atoms. Indeed, the ionic radius of sixfold coordinated $Bi^{3+}$ (1.03 Å) is larger than that of $Sb^{3+}$ (0.76 Å)[52]. Notably, both $Sb_2Te_3$ and $Bi_2Te_3$ compounds exhibit the most significant changes in distance within the fourth coordination shell, where the Sb-Sb interatomic distance (in the c-axis direction) expands from 4.64 to 4.70 Å, and the Bi-Bi distance increases from 4.76 to 4.81 Å as the temperature rises from 10 to 300 K [Figure 7A and C].

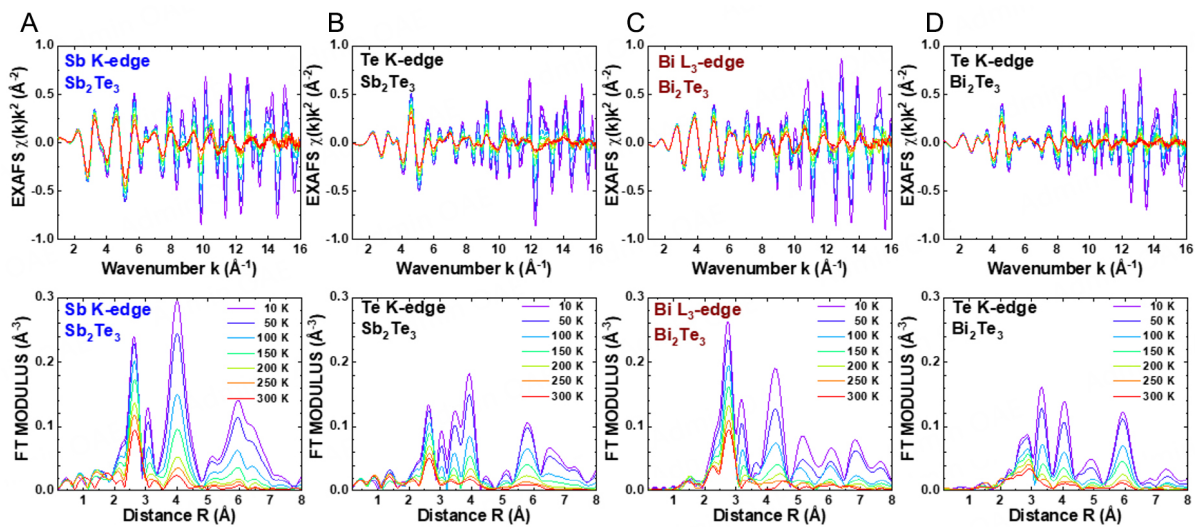

**Figure 6.** Temperature dependence of the experimental EXAFS spectra [$\chi(k)k^2$] and their Fourier transforms (FTs) for the Sb and Te K-edges in $Sb_2Te_3$ (A and B) and Bi $L_3$-edge and Te K-edge in $Bi_2Te_3$ (C and D). The FTs were calculated in the *k*-space range of 3.0-15.4 Å$^{-1}$, and only the FT modulus is shown for clarity.

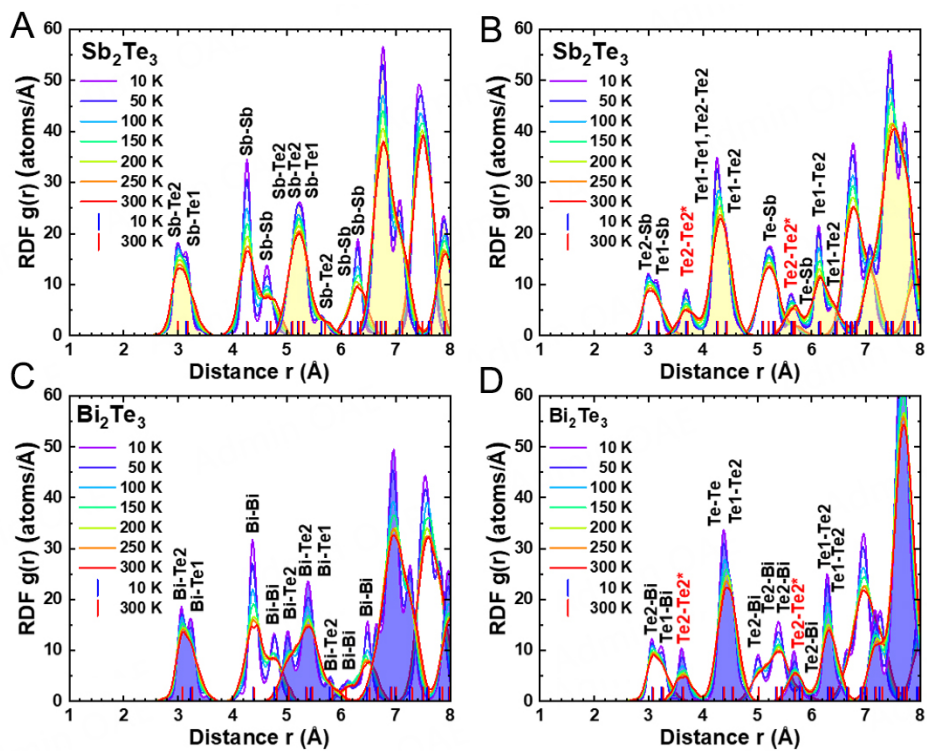

**Figure 7.** Temperature-dependent radial distribution functions (RDFs) *g*(*r*) calculated from the results of RMC/EA simulations for $Sb_2Te_3$ (A and B) and $Bi_2Te_3$ (C and D). Vertical lines show calculated average interatomic distances at 10 K (blue) and 300 K (red). The nearest atom pairs are indicated for clarity. The Te2-Te2 atom pairs from different layers are highlighted in red.

Temperature dependencies of the MSRD factors for the nearest Sb (Bi)-Te, Sb (Bi)-Sb (Bi), and Te-Te atom pairs in $Sb_2Te_3$ and $Bi_2Te_3$ are plotted in Figure 8, together with corresponding fits from the correlated Einstein model and estimated effective force constants. Note that data points at 300 K were excluded from the fitting procedure due to potential errors associated with the small amplitude of the experimental EXAFS spectra. The temperature dependence curves in both materials exhibit pronounced steepness that suggests relatively weak bonding between Sb, Bi, and Te atoms. Throughout all temperatures, the MSRD factor for the first coordination shell (M-Te2) consistently remains lower than that of the second coordination shell (M-Te1). Additionally, there is a notable distinction in the temperature dependencies of MSRD(M-Te2) and MSRD(M-Te1) [Figure 8D and G], arising from the fact that weak vdW forces connect two adjacent Te2 layers. At the same time, Te1 atoms are situated between two Sb (Bi) layers [Figure 8A]. The bond between M-Te2 is predominantly covalent, whereas the M-Te1 bond exhibits a combination of covalent and ionic characteristics[25].

In $Sb_2Te_3$, the strength of the Sb-Te2 bond [$f(r1_{Sb-Te2})$ = 24.2 ± 1.4 N/m] in the first coordination shell is notably larger than that of the Sb-Te1 bond [$f(r2_{Sb-Te1})$ = 14.8 ± 0.6 N/m] in the second shell. The corresponding Einstein temperatures 116.9 ± 3.4 K and 91.3 ± 1.8 K are smaller than those (151 ± 2.4 K and 129.1 ± 3.4 K) reported for microcrystalline $Sb_2Te_3$[44]. Notably, the Sb interactions with the Te2 from the second quintile exhibit a similar strength [$f(r5_{Sb-Te2})$ = 23.0 ± 1.8 N/m] as those within the first coordination shell. At the same time, the MSRD factor for the Sb-Te2 atom pair in the fifth coordination shell ($r5_{Sb-Te2}$) is the largest at 10 K due to increased static disorder [Figure 8A]. Within the layers of $Sb_2Te_3$, the interactions involving the nearest Sb-Sb and Te-Te atom pairs are comparable: $f(r3_{Sb-Sb})$ = 18.6 ± 0.7 N/m and $f(r4_{Te-Te})$ = 18.5 ± 0.7 N/m, respectively. Across the vdW gap, the Te2-Te2 bond strength [$f(r3_{Te2-Te2})$ = 17.4 ± 0.8 N/m] is slightly lower. The corresponding Einstein temperature for the third coordination shell (Sb-Sb) is 103.7 ± 1.9 K, which is very close to that of microcrystalline $Sb_2Te_3$ (109.7 ± 2.2 K)[44].

In $Bi_2Te_3$, the effective force constant for the first [$f(r1_{Bi-Te2})$ = 25.7 ± 1.4 N/m] and second [$f(r2_{Bi-Te1})$ = 14.5 ± 0.4 N/m] coordination shells are close to those observed in $Sb_2Te_3$ [$f(r1_{Sb-Te2})$ = 24.2 ± 1.4 N/m and $f(r2_{Sb-Te1})$ = 14.8 ± 0.6 N/m]. Within the layers of $Bi_2Te_3$, interactions involving the nearest Bi-Bi atom pairs [$f(r3_{Bi-Bi})$ = 17.4 ± 0.5 N/m] are close to those between Te-Te atom pairs [$f(r4_{Te-Te})$ = 18.1 ± 1.1 N/m]. The interactions between Te2 atoms across the vdW gap of $Bi_2Te_3$ [$f(r3_{Te2-Te2})$ = 16.1 ± 0.6 N/m] are lower. The corresponding Einstein temperatures for the first two coordination shells (106.7 ± 2.9 K and 80.3 ± 1.6 K) are smaller than those determined for microcrystalline $Bi_2Te_3$ (143.4 ± 2.3 K and 122.1 ± 2.7 K[44]). However, for the third coordination shell (Bi-Bi), the Einstein temperature of 76.5 ± 1.1 K coincides with that (77.7 ± 2.1 K) reported earlier[44].

Thus, we have demonstrated that the analysis of temperature-dependent X-ray absorption spectra based on atomistic simulations such as RMC method gives an access to local interatomic interactions described by effective force constants, which, in turn, affect how phonons propagate through the lattice. A lower force constant usually results in softer bonds and lower phonon frequencies, leading to lower $\kappa$, which is desirable for TEs[53]. According to interatomic force constants reported in [Figure 8D and G], the large differences in the temperature dependencies of MSRD(Sb/Bi-Te2) and MSRD(Sb/Bi-Te1) and related force constants indicate high anisotropy of the $\kappa$ in $Bi_2Te_3$ and $Sb_2Te_3$ in the directions along and across the QLs in their crystallographic structure[54]. Note also that while most force constants, reported in Figure 8D-I, have close values in the two compounds, $f(r5_{Sb/Bi-Te2})$ and $f(r6_{Sb/Bi-Te2})$ force constants are slightly smaller in $Bi_2Te_3$.

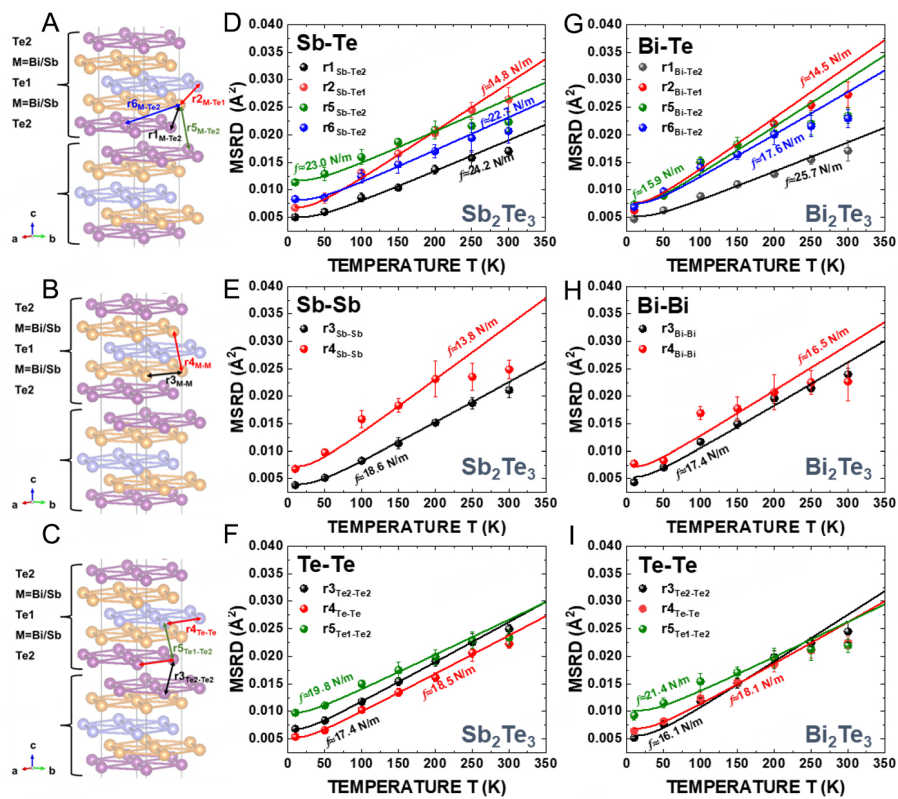

**Figure 8.** Temperature dependence of the MSRD factors, determined through RMC/EA calculations for nearest Sb (Bi)-Te, Sb(Bi)-Sb(Bi), and Te-Te atom pairs in $Sb_2Te_3$ (D-F) and $Bi_2Te_3$ (G-I). Corresponding atom pairs are highlighted in panels (A-C). The solid lines represent fits using the correlated Einstein model, with the exclusion of data points at 300 K. Corresponding effective force constants (*f*) are also indicated.

**Transport property evaluation on SPS-sintered pellets**

The TE transport data for the SPS-sintered $Bi_2Te_3$ and $Sb_2Te_3$ pellets are presented in Figure 9, in the temperature interval from 300 to 575 K. It is important to note that the data obtained for the binary TE phases of $Bi_2Te_3$ and $Sb_2Te_3$ synthesized in this work are compared only with the relevant binary phases (not ternary, quaternary, or secondary phase doped solid-solutions). The secondary phases of $Bi_2O_3$, $Sb_2O_3$, and $TeO_x$ have larger band gaps than the semiconducting phases[55-57]. Based on the XRPD and XPS data, these phases are more surface-dominant. Due to their low concentration within the main matrix, they are not expected to affect electronic transport significantly. However, due to the difference in their crystal structure, the oxide phases may act as phonon scattering centers, influencing thermal conductivity.

Figure 9A shows the σ for both $Bi_2Te_3$ and $Sb_2Te_3$, displaying a decreasing trend by increasing temperature, which is a typical behavior of metals or heavily doped semiconductors. The σ for $Sb_2Te_3$ changes from ~2,000 S/cm at 300 K to ~700 S/cm at 575 K, while it changes from ~1,750 S/cm at 300 K to ~1,000 S/cm at 575 K for $Bi_2Te_3$. A decrease in σ above 300 K typically indicates a rise in electron-phonon scattering. The σ of both samples is higher than those of other solution-derived samples and is also comparable with ingot alloy (on the level of 1,000 S/cm at 300 K)[58-60]. The high σ can be attributed to the layered structure and large-sized grains of both $Bi_2Te_3$ and $Sb_2Te_3$.

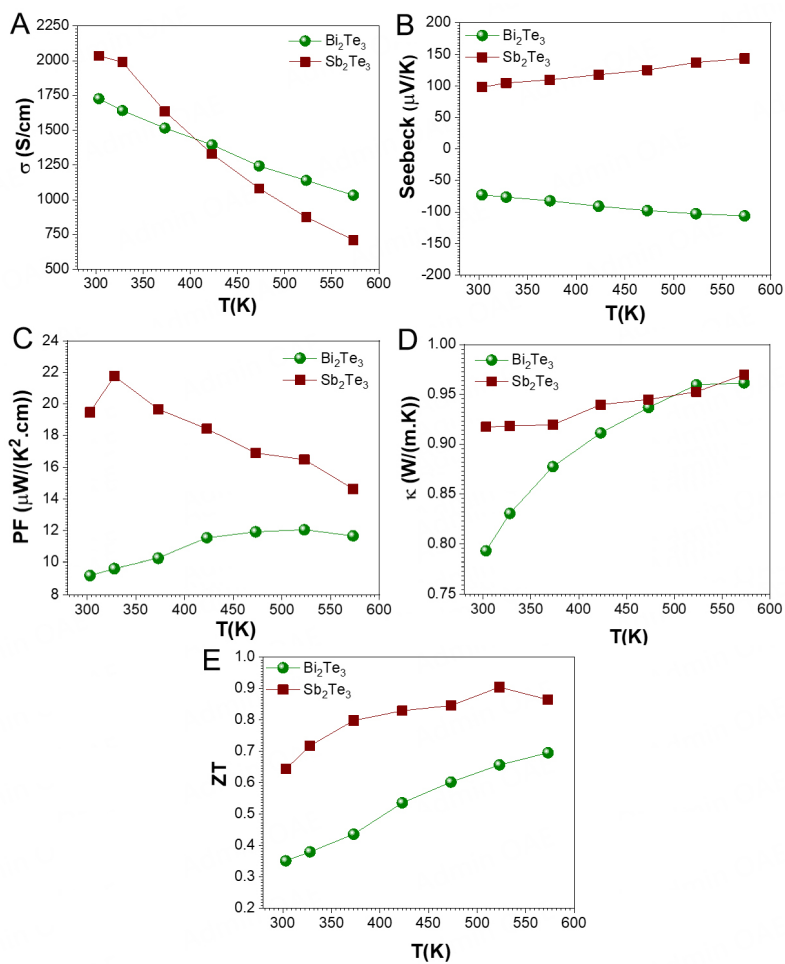

**Figure 9.** Temperature-dependent electronic and thermal transport properties of SPS sintered $Sb_2Te_3$ and $Bi_2Te_3$ pellets from nanopowders synthesized through MW-assisted thermolysis route; Electrical conductivity-σ (A), Seebeck coefficient-*S* (B), Power factor-PF (C), Thermal conductivity-$κ_{tot}$ (D), and the TE figure of merit ZT (E).

The *S* mainly demonstrates the type of transport in the TE materials and shows an opposite trend (in absolute value) to the σ. Figure 9B shows the obtained *S* values: a negative value for $Bi_2Te_3$ starting from -70 μV/K at 300 K and reaching -106 μV/K at 575 K; and a positive value for $Sb_2Te_3$, starting at 100 μV/K at 300 K and reaching to 145 μV/K at 575 K. The sign of *S* confirms characteristic n- and p-type transport in the synthesized $Bi_2Te_3$ and $Sb_2Te_3$, respectively. The *S* values obtained are in the same order, or slightly lower than other solution-derived samples, due to the high σ values obtained for these samples[59,60].

The power factor (PF) ($S^2σ$) was calculated for both $Bi_2Te_3$ and $Sb_2Te_3$ and is graphically presented in Figure 9C. The PF for $Bi_2Te_3$ at room temperature was estimated as 9 μW·$cm^{-1}$ $K^{-2}$, and the maximum value of 12 μW·$cm^{-1}$ $K^{-2}$ was reached at 523 K. For $Sb_2Te_3$, the room temperature value was 19 μW·$cm^{-1}$ $K^{-2}$, and the highest value was 22 μW·$cm^{-1}$ $K^{-2}$ around 328 K. The magnitude of PF for $Bi_2Te_3$ is slightly lower than the room temperature PF of samples prepared using other solution-chemical syntheses methods of polyol (16 μW·$cm^{-1}$ $K^{-2}$) and hydrothermal (24 μW·$cm^{-1}$ $K^{-2}$) routes[3]. The difference is mainly attributed to the high σ value, and the low *S* of samples presented in this work. The magnitude of PF obtained for $Bi_2Te_3$ and $Sb_2Te_3$ is, however, higher than many other wet-chemically synthesized TE[61-63] materials with a maximum PF of 1-9 μW·$cm^{-1}$ $K^{-2}$.

The $\kappa_{tot}$ values were determined by measuring the thermal diffusivity, mass density, and heat capacity of samples. Figure 9D illustrates the $\kappa_{tot}$ of SPS-consolidated samples. The microstructure analysis of the sintered samples [Figure 4] reveals smaller grains and, consequently, a higher density of grain boundaries in $Bi_2Te_3$, resulting in a lower $\kappa_{tot}$. The overall upward trend in $\kappa_{tot}$ for both the samples is attributed to the bipolar effect in materials with a narrow band gap, and minority charge carriers are easily excited as the temperature increases, contributing to enhanced heat conduction and introducing increased $\kappa_{tot}$ to the samples[64,65]. Across the entire temperature range, it is evident that the $\kappa_{tot}$ for both samples remains consistently below 0.96 W/m·K. This observation underscores the significant impact of nanostructuring, anisotropy, and texturing on the phonon scattering mechanism. In an earlier work on wet-chemically synthesized $Bi_{2-x}Sb_xTe_3$, room temperature $\kappa_{tot}$ values in the range 1-1.9 W/m·K have been reported for materials with similar composition[60]. Furthermore, studies performed on various nanostructures of $Bi_2Te_3$/$Sb_2Te_3$ revealed room temperature $\kappa_{tot}$ values of 1.47 and 1.81 W/m·K[66]. Our $\kappa_{tot}$ values are about 10% to 50% lower than these values, which may be associated with the increase in anisotropy due to the large layered microstructure of the samples.

The ZT has been calculated for both samples based on their electrical and thermal transport data, and the results are presented in Figure 9E. At room temperature, the ZT value for the $Sb_2Te_3$ sample is 0.65, reaching its highest value of 0.9 at 523 K. The ZT for $Bi_2Te_3$ exhibits a linear increase with increasing temperature, starting from about 0.35 at room temperature, reaching its peak value of 0.7 at 573 K. A comparison of ZT values for $Bi_2Te_3$ and $Sb_2Te_3$ samples in this work with the earlier reports is presented in Figure 10. $Bi_2Te_3$ synthesized through thermolysis showed a ZT of 0.62 (at 400 K)[67]. A ZT value of 0.84 (at 423 K) was reported for $Bi_2Te_3$ synthesized through the reduction of mechanically milled oxide powders[68], and 0.36 (at 325 K) for the solvothermally synthesized sample[60]. We have earlier reported ZT values of 0.9 (at 373 K) and 1.03 (at 473 K) for single-phase sintered $Bi_2Te_3$ synthesized using MW-assisted polyol and hydrothermal techniques[3]. Furthermore, we reported earlier a ZT value of 1.37 (at 523 K) for single-phase sintered $Sb_2Te_3$ samples synthesized via the MW-assisted polyol route[49]. An earlier work on $Sb_2Te_3$ prepared using MW-assisted thermolysis (~15 min) showed a ZT of 0.96 (at 423 K)[69]. A ZT of 0.55 was reported[70] for hydrothermally synthesized $Sb_2Te_3$. Meanwhile, MW-assisted polyol synthesis yielded a ZT of 0.58 (at 420 K)[36]. At last, $Sb_2Te_3$ synthesized through solvothermal method yielded a ZT of 0.13 (at 423 K)[71] As can be seen from these values, the highest ZT achieved for these materials (and the temperature at which the reported ZT is reached) is rather scattered for the TE materials with the same composition, which can be attributed to the different synthetic methods employed, which substantially influence the resulting microstructure, surface chemistry, and the type and content of impurity phases. These variables significantly influence the overall TE performance of the materials. One important advantage of the presented synthesis route is the easy dispersibility of the as-synthesized materials in polymer matrices, enabling very homogeneous hybrid material design[32]. With substantially reduced time and carbon footprint, the developed synthetic method provides a promising sustainable approach for large-scale synthesis of high-purity nanostructured TE materials with promising performance, at a yield greater than 95%. It is important to note that the maximum ZT temperature of samples presented here shifted significantly to the high-temperature region, highlighting their potential for power generation applications.

## CONCLUSIONS

A rapid, energy-efficient, and scalable solution-based synthesis method was developed, using MW-assisted thermolysis process - at 220 °C within 6 min, obtaining successfully bismuth and antimony telluride ($Bi_2Te_3$, $Sb_2Te_3$) as n- and p-type TE materials. As-made materials were characterized using various techniques, comprising XRPD, SEM, TEM, XAS, and XPS. XRPD confirmed a rhombohedral layered crystal structure, while SEM revealed hexagonal platelet-shaped nanoparticles with lateral dimensions in the range of 20-4,000 nm. XPS analysis revealed the presence of $Sb^{3+}$, $Bi^{3+}$, and $Te^{4+}$, attributed to surface-bound oxide layers in the form of $Bi_2O_x$, $Sb_2O_3$, and $TeO_2$. Detailed investigation of the local atomic structure in the synthesized $Bi_2Te_3$ and $Sb_2Te_3$ powder samples was conducted through synchrotron radiation XAS experiments. RDFs around absorbing atoms were reconstructed using RMC simulations, and effective force constants for the nearest and distant coordination shells were subsequently determined from the temperature dependencies of the MSRDs of the Sb(Bi)-Te, Sb(Bi)-Sb(Bi), and Te-Te atom pairs. The observed differences in the effective force constants support high anisotropy of the thermal conductivity in $Bi_2Te_3$ and $Sb_2Te_3$ in the directions along and across the QLs in their crystallographic structure. The as-made materials were consolidated via SPS, upon which texturing along the c-axis was observed in both samples. Electrical and thermal transport properties were determined on the sintered pellets by measuring electrical resistance and the *S*, while the thermal diffusivity was measured using the LFA system. The negative sign of the *S* identifies the n-type character of $Bi_2Te_3$, while the positive sign indicates the p-type character of $Sb_2Te_3$. Promising $\sigma$ values were obtained for both samples, though Seebeck values were slightly lower than those reported in earlier works. It was found that due to the effective phonon scattering, attributed to nanostructuring, the sintered TE materials exhibited low thermal conductivity, achieving the highest ZT values of 0.7 (at 573 K) and 0.9 (at 523 K) for n-type $Bi_2Te_3$ and p-type $Sb_2Te_3$, respectively. These results surpass those reported for these materials synthesized through many other wet-chemical synthetic routes. The TE performance of the synthesized materials shows good batch-to-batch reproducibility, where the highest ZT has shifted significantly to the high-temperature region, highlighting their potential for power generation applications. The presented method is truly scalable and can easily be tailored for one-pot synthesis of ternary and quaternary TE compositions in one-pot. The scalable, energy- and time-efficient synthetic method developed, along with the demonstration of its potential for TE materials, creates opportunities for broader use of these strategic materials while minimizing environmental impact.

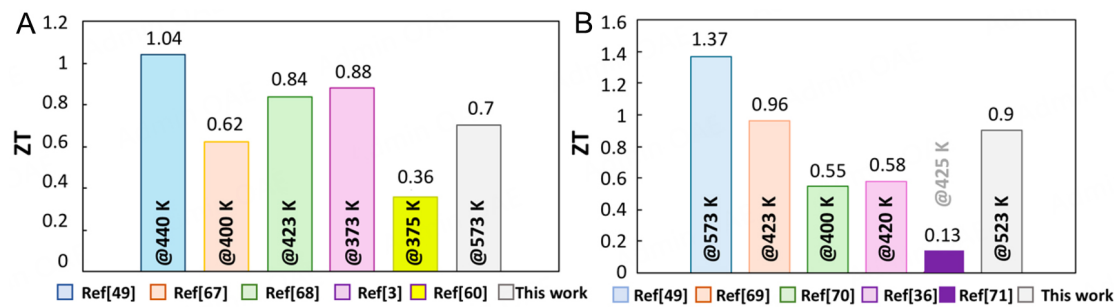

**Figure 10.** Comparison of ZT values of sintered (A) $Bi_2Te_3$ and (B) $Sb_2Te_3$ samples with earlier reports on binary phases synthesized through different chemical routes. The temperature at which the reported ZT value is reached for the different samples is displayed on the histograms.

## DECLARATIONS
**Acknowledgments**
Hamawandi, B., Pudza, I., and Pudzs, K. thank the Latvian Council of Science for support through project No. lzp-2023/1-0528. Toprak, M. S. acknowledges funding from the European Union's Horizon 2020 research and innovation program under grant agreement No. 863222. Ballikaya, S. acknowledges support from the Scientific and Technological Research Council of Turkey (TUBITAK, 119N120). We acknowledge DESY (Hamburg, Germany), a member of the Helmholtz Association HGF, for providing experimental facilities. The experiment at the DESY PETRA III synchrotron was performed within proposal No. I-20220381 EC. Institute of Solid State Physics, University of Latvia, as the Center of Excellence, has received funding from the European Union's Horizon 2020 Framework Programme H2020-WIDESPREAD-01-2016-2017-TeamingPhase2 under grant agreement No. 739508, project CAMART². Toprak, M. S. also acknowledges the financial support from Olle Engkvist Foundation (SOEB, 190-0315) for establishing MW synthesis facilities.

**Authors' contributions**
Conception and design of the study, methodology, writing - original draft, writing - review & editing, supervision, funding acquisition: Toprak, M. S.
Conception, methodology, writing - original draft, writing - review & editing: Hamawandi, B.
Investigation, data analysis and interpretation, visualization, writing - original draft, writing - review & editing, funding acquisition: Pudza, I.
Investigation: Pudzs, K.
Investigation, methodology, writing - original draft, writing - review & editing, supervision: Kuzmin, A.
Resources: Welter, E.
Investigation, methodology, data analysis and interpretation: Ballikaya, S.
Investigation, methodology: Parsa, P.; Szukiewicz, R.; Kuchowicz, M.

**Availability of data and materials**
Data will be available from the corresponding author upon reasonable request.

**Financial support and sponsorship**
This work was supported by the Latvian Council of Science (project No. lzp-2023/1-0528), the European Union's Horizon 2020 research and innovation program under grant agreement No 863222 (UncorrelaTEd), Scientific and Technological Research Council of Turkey (TUBITAK, 119N120), and the European Union's Horizon 2020 Framework Programme H2020-WIDESPREAD-01-2016-2017-TeamingPhase2 under grant agreement No. 739508, project CAMART². The funding bodies had no role in the experiment design, collection, analysis and interpretation of data, and writing of the manuscript.

**Conflicts of interest**
Toprak, M. S. is an Editorial Board Member of the journal *Energy Materials*. Toprak, M. S. was not involved in any steps of editorial processing, notably including reviewers' selection, manuscript handling and decision making, while the other authors have declared that they have no conflicts of interest.

**Ethical approval and consent to participate**
Not applicable.

**Consent for publication**
Not applicable.